\documentclass[final]{article}

\usepackage{arxiv}

\usepackage[utf8]{inputenc} 
\usepackage[T1]{fontenc}    
\usepackage{url}            
\usepackage{booktabs}       
\usepackage{amsfonts}       
\usepackage{nicefrac}       
\usepackage{microtype}      
\usepackage{lipsum}
\usepackage[range-phrase=--,range-units=single,separate-uncertainty,multi-part-units = single,detect-weight=true,]{siunitx}
\usepackage{textcomp} 

\usepackage{amsmath}
\usepackage{amssymb}
\usepackage{graphicx} 
\usepackage{dcolumn} 
\usepackage{bm} 
\usepackage{blindtext}
\usepackage{xcolor}
\usepackage[title]{appendix}
\usepackage[superscript]{cite}
\usepackage{stfloats} 
\usepackage{hyperref}
\usepackage{cleveref}

\newcommand{\td}[2]{%
  \frac{\mathrm{d}#1}{\mathrm{d}#2}%
  }
\newcommand{\ttd}[2]{%
  \frac{\mathrm{d}^{2}#1}{\mathrm{d}#2^{2}}%
  }

\newcommand{\tdi}[2]{%
  \mathrm{d}#1/\mathrm{d}#2%
  }
\newcommand{\ttdi}[2]{%
  \mathrm{d}^{2}#1/\mathrm{d}#2^{2}%
  }

\crefname{equation}{Eq.}{Eqs.}
\Crefname{equation}{Equation}{Equations}

\crefname{table}{Table}{Tables}
\Crefname{table}{Table}{Tables}

\crefname{figure}{Fig.}{Figs.}
\Crefname{figure}{Figure}{Figures}

\crefname{section}{Sec.}{Secs.}
\Crefname{section}{Section}{Sections}

\crefname{chapter}{Chap.}{Chaps.}
\Crefname{section}{Section}{Sections}

\title{Acoustic Source Localization with the Angular Spectrum Approach in Continuously Stratified Media}

\author{
  Scott Schoen Jr\\\
  Woodruff School of Mechanical Engineering\\
  Georgia Institute of Technology \\
  Atlanta, GA USA \\
  \texttt{scottschoenjr@gatech.edu} \\
   \And
 Costas D. Arvanitis \\
  Woodruff School of Mechanical Engineering and Coulter Department of Biomedical Engineering\\
  Georgia Institute of Technology and Emory University\\
  Atlanta, GA, USA \\
  \texttt{costas.arvanitis@gatech.edu} 
}

\begin{document}
\maketitle
\frenchspacing

\vspace{-5mm}
\begin{center}
  \begin{minipage}[]{0.685\textwidth}
    \small\textbf{%
    The following article has been accepted by \textit{JASA Express Letters}. 
    After publication, it will be found \href{http://asa.scitation.org/journal/jel}{here}.
    }
  \end{minipage}
\end{center}
\vspace{5mm}

\begin{abstract}
The angular spectrum approach (ASA)---a fast, frequency domain method for calculation of the acoustic field---enables passive source localization and modeling forward propagation in homogeneous media with high computational efficiency.
Here we show that, if the medium is continuously stratified,  a first-order analytical solution may be obtained for the field at arbitrary depth. 
Our simulations show that the stratified ASA solution enables accurate source localization as compared to the uncorrected ASA (error from $1.2\pm0.3$ to $0.49\pm0.3$ wavelengths) at scalings relevant to biomedical ($kL \sim 500$, where $L$ is the length of the measurement aperture), underwater ($kL \sim 800$), and atmospheric ($kL \sim 10$) acoustic applications. 
Overall the total computation was on the order milliseconds on standard hardware (\SI{225\pm84}{ms}, compared with \SI{78\pm63}{ms} for the homogeneous ASA formulation over all cases). 
Collectively, the results suggest the proposed ASA phase correction enables efficient and accurate method for source localization in continuously stratified environments.
\end{abstract}


\section{Introduction}
\label{sec:Introduction}
The angular spectrum approach~(ASA) is a method of  solving the wave equation in the spatial frequency domain.\cite{goodman_introduction_1996}
It may be considered a decomposition of the harmonic field into plane waves that travel with a continuous spectrum of directions, with appropriate phases and amplitudes such that their summation yields fields of arbitrary spatial distribution.\cite{booker_concept_1950}
Plane waves are mathematically convenient, as the propagation through space of each component may be modeled with a simple phase delay.
Coupled with fast Fourier transform (FFT) algorithms, the method enables exceptionally efficient computations and is naturally suited to real-time applications with relatively narrow-band sources.
Of particular interest for the ASA here is the recovery of the volumetric wave field from the measured pressure on a planar surface (i.e., boundary condition), a process known as acoustical holography.\cite{williams_fourier_1999}
From the ASA reconstruction, a passive acoustic map~(PAM) may be formed at specific frequencies, and thus sources at these frequencies to be localized as intensity peaks in the PAM. 

In practical applications, this surface measurement is made by a series of sensors (e.g., an ultrasonic imaging transducer or linear array of hydrophones) which sample discretely the pressure field due to sources in the region interest. 
Such source localization from PAMs is a problem of significant interest in biomedical,\cite{gyongy_passive_2010,haworth_passive_2012,arvanitis_passive_2017} underwater,\cite{maranda_efficient_1989,chen_novel_2010,chi_fast_2016} and aeroacoustic\cite{lee_impedance_1986,tien_pham_adaptive_1996} applications.
However, in such applications the propagation environment typically varies as a function of the vertical direction.
While recent work\cite{schoen_jr_heterogeneous_2020} has shown that heterogeneity can be accounted for with the ASA, an analytical solution might ensure high computational efficiency for large imaging domains to reduce errors in the source localization.

Herein we derive a first order analytical correction for the ASA and for the case of a stratified medium. 
Then, through simulated acoustic propagation, we demonstrate corrected sub-wavelength localization in stratified environents at scales relevant to biological, atmospheric, and underwater acoustics.
Finally, we consider the point source localization errors, and the computational cost of the different algorithm permutations.

\section{Methods}

\subsection{Angular Spectrum Approach for Stratified Media}
\label{sec:AngularSpectrum}
The angular spectrum $P(k_{x}, k_{y}, z)$ of a monochromatic ($\propto -i\omega t$) field $\tilde{p}(x, y, z)$ defined by 
\begin{align}
  P 
  = 
  \mathcal{F}_{k}[\,\tilde{p}\,] 
  \equiv 
  \iint_{-\infty}^{\infty}{ \tilde{p}(x, y, z)e^{-i\left(k_{x}x + k_{y}y \right)}\,\mathrm{d}x\,\mathrm{d}y}\,.
  \label{eqn:AngularSpectrumDefinition}
\end{align}
Taking the 2D Fourier transform to the homogeneous Helmholtz equation $(\nabla^{2} + \omega^{2}/c_{0}^{2})\tilde{p} = 0$ with a constant speed of sound $c_{0}$ gives an ordinary differential equation for $P$
\begin{align}
  \ttd{P}{z} + k_{z}^{2}P = 0\,,
  \label{eqn:TransformedAsOde}
\end{align}
where $k_{z}^{2} = (\omega/c_{0})^{2} - k_{x}^{2} - k_{y}^{2}$. 
Knowledge of the boundary condition $P_{0}$ at $z = 0$, and the assumption that there are no backward-travelling waves, then \cref{eqn:TransformedAsOde} has the solution
\begin{align}
  P &= P_{0}e^{ik_{z}z}\,.
  \label{eqn:AngularSpectrumTransferFunction}
\end{align}
The acoustic field in any plane may then be reconstructed by evaluation of the inverse transform of $P$ given by \cref{eqn:AngularSpectrumTransferFunction}.

Suppose now that the sound speed has weak spatial dependence (i.e., that the sound speed changes over scales that are large compared with the wavelength), such that $c_{0} \rightarrow c(\boldsymbol{r})$ in the Helmholtz equation.
Then, with the definitions $\mu = c_{0}^{2}/c^{2}(\boldsymbol{r})$ and $\lambda = (1 - \mu)\omega^{2}/c_{0}^{2}$ for a constant reference sound speed $c_{0}$, it can be shown that the governing equation is
\begin{align}
 \ttd{P}{z} + k_{z}^{2}P 
  &= 
  \Lambda * P\,.
  \label{eqn:InhomogeneousGoverningEquation}
\end{align}
Here $\Lambda = \mathcal{F}_{k}\left[\,\lambda\,\right]$, and $*$ represents the 2D convolution over the wavenumber components $k_{x}$ and $k_{y}$.
If $c$ varies only with the axial coordinate $z$, the convolution may be evaluated to give
\begin{align}
  \ttd{P}{z} + k_{z}^{2}P &= \lambda\,P\,.
  \label{eqn:DiffEqInAngularSpectrum}
\end{align}
Assume a Wentzel–Kramers–Brillouin (WKB)--type solution\cite{morse_methods_1953,agrawal_angular_1975} of the form
\begin{align}
  P &= A( k_{x}, k_{y}, z )\,e^{ik_{z}z}\,,
  \label{eqn:StratefiedMediumAssumedSolution}
\end{align}
where $A$ is a complex amplitude.
Substitution of \cref{eqn:StratefiedMediumAssumedSolution} into \cref{eqn:DiffEqInAngularSpectrum} and evaluation of the derivatives yields
\begin{align}
  \ttd{A}{z} + 2ik_{z}\td{A}{z} - \lambda A &= 0\,.
  \label{eqn:StratifiedMediumOdeInAmplitude}
\end{align}
To first order, the first term in \cref{eqn:StratifiedMediumOdeInAmplitude} can be neglected to obtain a first-degree ODE for $A$, which may be integrated directly to give
\begin{align}
  A &= A_{0}\operatorname{exp}{\left( \frac{1}{2ik_{z}}\int_{0}^{z}{\lambda(z)\,\mathrm{d}z'} \right)}\,.
  \label{eqn:LinearizedAngularSpectrumAmplitudeStratifiedIntermediate}
\end{align}
Neglecting the higher order terms requires that changes in sound speed occur on scales that are very large compared to the wavelength ($|(\tdi{c}{z})\Delta z/c|$), and that the relative magnitude of these changes is not too large ($|c/c_{0}| \sim 1$, see Supplementary Material).
Application of the boundary condition (i.e., the Fourier transform of the measured field at $z = 0$) gives
\begin{align}
  A &= P_{0}\operatorname{exp}{\left( \frac{1}{2ik_{z}}\int_{0}^{z}{\lambda(z')\,\mathrm{d}z'} \right)}\,.
  \label{eqn:AngularSpectrumComplexCoefficient}
\end{align}
The angular spectrum at arbitrary $z$ is then
\begin{align}
  P 
  &= 
  \left[ P_{0}\operatorname{exp}{\left( \frac{1}{2ik_{z}}\int_{0}^{z}{\lambda(z')\,\mathrm{d}z'} \right)} \right]e^{ik_{z}z}
  \nonumber \\
  &= 
  P_{0}\operatorname{exp}{\left[ i\left(k_{z}z - \frac{k_{0}^{2}}{2k_{z}}\int_{0}^{z}{1 - \mu(z')\,\mathrm{d}z'}\right) \right]} \,.
  \label{eqn:LinearizedAngularSpectrumAmplitudeStratified}
\end{align}

\begin{figure}
    \centering
    \includegraphics[width=\textwidth]{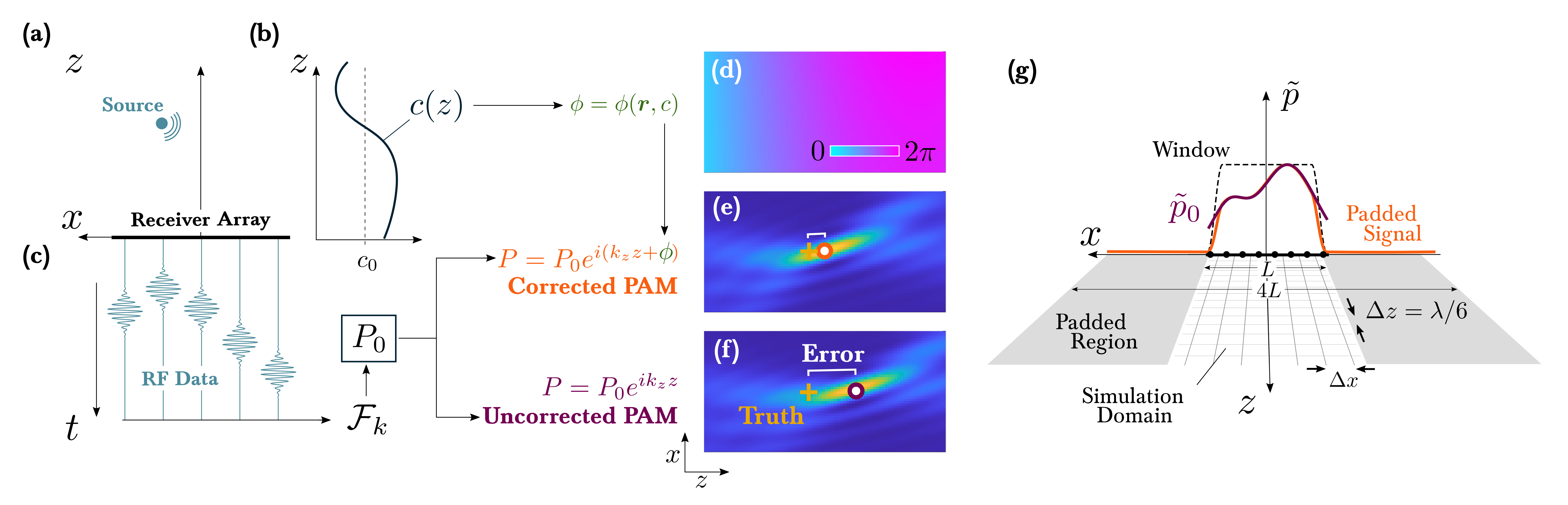}
    \caption{Aberration correction for Passive Acoustic Mapping %
    \textbf{(a)}~Emissions from an acoustic source are recorded with a receiver array at the bottom of the half-space. %
    \textbf{(b)}~The medium has a sound speed that varies in the $z$-direction.  %
    \textbf{(c)}~The time series data collected by the array elements are transformed into the spatial frequency domain, to obtain the ASA initial condition $P_{0}$. %
    \textbf{(d)}~The varying sound speed $c(z)$ is used to compute a spatially dependent phase correction $\phi$. %
    \textbf{(e)}~Use of the phase-corrected ASA [\cref{eqn:LinearizedAngularSpectrumAmplitudeStratified}] enables improved localization of the source compared to %
    \textbf{(f)}~conventional ASA beamforming [\cref{eqn:AngularSpectrumTransferFunction}]. %
    \textbf{(g)}~For all computations, the measured signals were windowed and zero-padded.}
    \label{fig:PhaseCorrectionConcept}
\end{figure}

\subsubsection{First-Order Restrictions}
Neglecting the second-order term in \cref{eqn:StratifiedMediumOdeInAmplitude}, yields the solution given by \cref{eqn:AngularSpectrumComplexCoefficient}. 
This assumption requires that $|\ttdi{A}{z}|$ is small compared to both $|2ik_{z}(\tdi{A}{z})|$ and $|\lambda A|$ [recall that here $\lambda \equiv k_{0}^{2}(1 - \mu)$ rather than the wavelength].
Evaluation of the derivatives in \cref{eqn:AngularSpectrumComplexCoefficient} gives the following condition
\begin{align}
  \left| \ttd{A}{z} \right| 
  =
  \left| \frac{\tdi{\lambda}{z}}{2ik_{z}} - \frac{\lambda^{2}}{4k_{z}^{2}}\right| |A|
  \implies
  \left| \frac{\tdi{\lambda}{z}}{2ik_{z}\lambda} - \frac{\lambda}{4k_{z}^{2}}\right|
  \leq
  \left| \frac{\tdi{\lambda}{z}}{2ik_{z}\lambda}\right|
  +
  \left|\frac{\lambda}{4k_{z}^{2}}\right|
  \ll
  1
  \label{eqn:ShortWavelengthCriterion}
\end{align}
The second term in the final expression of \cref{eqn:ShortWavelengthCriterion} dictates that  $|(k_{0}^{2}/k_{z}^{2})(1 - \mu)|$ is negligible.
Therefore when $k_{0}\sim k_{z}$  (paraxial approximation), this requirement is that $\mu \simeq 1 - c'/2c_{0} \sim 1$, i.e., that the relative magnitude of the sound speed changes should be small.
From the definition $\lambda = k_{0}^{2}(1 - \mu)$.
Thus the first term of the final expression in \cref{eqn:ShortWavelengthCriterion} requires that
\begin{align}
    \left| \frac{\tdi{\lambda}{z}}{2ik_{z}\lambda} \right|
    &=
    \left| \frac{k_{0}^{2}c_{0}^{2}\left[ \left(1/c^{3}\right)\tdi{c}{z}\right]}{2k_{z}k_{0}^{2}(1 - \mu)} \right|
    =
    \left| 
      \frac{1}{2k_{z}}
      \,
      \frac{\tdi{c}{z}}{c}
      \,
      \frac{\mu}{(1 - \mu)} 
     \right|
     \lesssim
     \left|
      \frac{\tdi{c}{z}}{\omega}
      \,
      \frac{\mu}{(1 - \mu)} 
     \right|\,.
     \label{eqn:SlowChangeCriterion}
\end{align}
 However, since it was required that $\mu \sim 1$, it must be true that $\tdi{c}{z}/\omega$, i.e., the sound speed change over a wavelength, is an order of magnitude smaller than $(c^{2}/c_{0}^{2} - 1)^{-1}$ for \cref{eqn:SlowChangeCriterion} to be true.
 Therefore, the high frequency requirement is significantly more important than the condition that the relative changes in the sound speed are small.

\subsubsection{Interpretation of Result}
\Cref{eqn:LinearizedAngularSpectrumAmplitudeStratified} represents an additional phase delay $\phi$ to the homogeneous medium case given by
\begin{align}
  \phi =\frac{k_{0}^{2}}{2k_{z}}\int_{0}^{z}{1 - \mu(z')\,\mathrm{d}z'}\,.
  \label{eqn:LinearizedAngularSpectrumAmplitudeStratifiedPhaseDelay}
\end{align}
Note that for a homogeneous medium, then $\mu = 1$, and the uniform case [\cref{eqn:AngularSpectrumTransferFunction}] is recovered.
\Cref{eqn:LinearizedAngularSpectrumAmplitudeStratifiedPhaseDelay} may be thought of as accumulation of phase shifts incurred as the wave travels through an infinitesimal width $\mathrm{d}z$, i.e.,
\begin{align}
  \phi = \int{\mathrm{d}\phi}
  \hspace{2mm}
  \implies 
  \hspace{2mm}
  \mathrm{d}\phi = \frac{k_{0}^{2}}{2k_{z}}\left( 1 - \mu\right)\,\mathrm{d}z\,.
  \label{eqn:LinearizedAngularSpectrumAmplitudeStratifiedPhaseDelayIncrement}
\end{align}
Since it was required that $c'/c_{0}$ is small, $\mu(z) = (1 + c'/c_{0})^{-2}$ can be expanded so that \cref{eqn:LinearizedAngularSpectrumAmplitudeStratifiedPhaseDelayIncrement} becomes
\begin{align}
  \mathrm{d}\phi &\simeq \frac{k_{0}^{2}}{2k_{z}}\left[ 1 - \left( 1 - 2\frac{c'}{c_{0}} \right)\right]\,\mathrm{d}z 
  \nonumber \\
  &\simeq \frac{k_{0}^{2}}{2k_{z}}\left( 2\frac{c'}{c_{0}} \right)\,\mathrm{d}z 
  =
  \left( \frac{c'}{c_{0}}k_{0} \right)\left( \frac{k_{0}}{k_{z}}\,\mathrm{d}z \right)\,.
\end{align}
The term $(c'/c_{0})k_{0}$ has the form of an effective wavenumber, accounting for the dilation of contraction of the wavelength due to the difference in sound speed from $c_{0}$.
The second term $(k_{0}/k_{z})\,\mathrm{d}z$ is the distance between the two infinitesimally separated planes for a plane wave traveling with propagating wavenumber $k_{z}$.
The extra phase then has a the familiar form $\phi \sim k_{\mathrm{eff}}d$.

\subsubsection{Comparison of Results}
We have shown in a previous work\cite{schoen_jr_heterogeneous_2020} that a numerical marching scheme may be applied to obtain an approximate solution to \cref{eqn:InhomogeneousGoverningEquation} in the case of general heterogeneity
\begin{align}
  P^{n + 1} 
  \approx 
  P^{n}e^{ik_{z}\Delta z}
  +
  \frac{e^{ik_{z}\Delta z}}{2ik_{z}}\,\left( P^{n} * \Lambda  \right) \times \Delta z\,,
  \label{eqn:AngularSpectrumMarchingAlgorithm}
\end{align}
where $P^{n} = P(k_{x}, k_{y}, n\Delta z)$.
To compare with the analytical result in the case of a stratified medium, \cref{eqn:LinearizedAngularSpectrumAmplitudeStratified} can be re-written with an expansion of the exponential term (see below) to give
\begin{align}
  P 
  &\simeq
  P_{0}e^{ik_{z}z}\left[%
    1 + \frac{1}{2ik_{z}}\int_{0}^{z}{\lambda(z')\,\mathrm{d}z'} 
    +
    \dots
  \right]\,.
  \label{eqn:ExponentialExpansion}
\end{align}
Because evaluation of \cref{eqn:AngularSpectrumMarchingAlgorithm} requires that $\Delta z$ is small compared to the wavelength,\cite{schoen_jr_heterogeneous_2020} the integral in \cref{eqn:ExponentialExpansion} may be approximated as a Riemann sum between each successive position $z$
\begin{align}
  \int_{z_{n}}^{z_{n} + \Delta z}{\lambda(z')\,\mathrm{d}z'}
  \simeq
  \lambda(z_{n})\times\Delta z
  +
  \mathcal{O}\left[ (\Delta z)^{2} \right]\,.
  \label{eqn:IntegralExpansion}
\end{align}
Use of \cref{eqn:IntegralExpansion} and retention of first order terms gives
\begin{align}
  P^{n + 1} \simeq P^{n}e^{ik_{z}\Delta z}
  + 
  \frac{e^{ik_{z}\Delta z}}{2ik_{z}}\,P^{n}\lambda(z)\,\Delta z
  +
  \mathcal{O}\left[ (\Delta z)^{2} \right]
  \,.
  \label{eqn:MarchingAlgorithmStratified}
\end{align}
In the stratified medium case, $\Lambda * P = \lambda P$, so that \cref{eqn:AngularSpectrumMarchingAlgorithm} agrees with \cref{eqn:MarchingAlgorithmStratified} to $\mathcal{O}\left[ (\Delta z)^{2} \right]$.
This is expected as the first-order solution of \cref{eqn:StratifiedMediumOdeInAmplitude} was used.

Use of the truncated expansion in \cref{eqn:ExponentialExpansion} requires that
\begin{align}
  \frac{1}{4}\left(\frac{k_{0}}{k_{z}}\right)^{2}\left[ k_{0}\int_{0}^{z}{(1 - \mu)\,\mathrm{d}z} \right]^{2} \ll 1\,.
  \label{eqn:TruncationCondition}
\end{align}
In the far field (e.g., for PAM), the paraxial approximation dictates that first term is of order 1.
\Cref{eqn:TruncationCondition} is true then if $\mu \equiv (c_{0}/c)^{2} \approx 1$, i.e., for relatively weak heterogeneity, which was assumed for use of the first-order solution of \cref{eqn:StratifiedMediumOdeInAmplitude}.

\subsection{Simulations and Numerical Implementation}
\label{sec:Simulations}
To determine the improvement in source localization, point sources were simulated in a half-space above a virtual receiving array in k-Wave;\cite{treeby_k-wave:_2010} see  \cref{fig:PhaseCorrectionConcept}(a).
For computational efficiency, simulations were performed in 2D, and thus in the simulations and reconstructions, $y = k_{y} = 0$.
The PAM reconstruction routines were written in MATLAB and computation times are reported for a standard desktop computer (Intel Core i7, four cores at 2.8~GHz and 16~GB memory); no parallel or graphical processing techniques were employed.
Sources were simulated as Gaussian pulses with \SI{5}{\percent} bandwidth.
The resulting pressure was measured by a virtual linear array, the acoustic data were then beamformed to compute $P(k_{x}, z)$ with \cref{eqn:AngularSpectrumTransferFunction} or \cref{eqn:LinearizedAngularSpectrumAmplitudeStratified}, for the uncorrected and corrected cases respectively.
Finally the intensity field was then calculated from
\begin{align}
  I(x, z) = \sum_{\omega}{\left\| \mathcal{F}_{k}^{-1}\left[ P(k_{x}, z)\right] \right\|^{2}}\,.
  \label{eqn:IntensityMapEquation}
\end{align}
Maps were formed over a frequency range $\omega$ of three frequency bins, centered at the source frequency.
The reconstructed source location $(x_{r}, z_{r})$ was taken to be the position of peak intensity of $I$, and the error was defined relative to the known source position $(x_{\mathrm{true}}, z_{\mathrm{true}})$ as
\begin{align}
  \epsilon = \sqrt{\epsilon_{x}^{2} + \epsilon_{z}^{2}} = \sqrt{(x_{r} - x_{\mathrm{true}})^{2} + (z_{r} - z_{\mathrm{true}})^{2}}\,.
\end{align}

To prevent spatial aliasing, all initial spectra were zero padded such that their computational spatial extent was four times larger than their physical extent. 
As the medium is stratified, the material properties are known for all $x$, and thus the computational domain may be extended arbitrarily from the simulation domain.
The measured spectra $\tilde{p}_{0}$ were windowed with a Tukey window with cosine fraction $R = 0.25$ to taper the high spatial frequency components that would be introduced by the discontinuity caused by the zero-padding.\cite{williams_fourier_1999} 
All reconstructions were performed with data from the virtual sensors spaced at $\Delta x$, and at depth steps $\Delta z = f_{0}/6c_{0}$ (i.e., 1/6 of the reference wavelength); see \cref{fig:PhaseCorrectionConcept}(g).
The discrete integration in \cref{eqn:LinearizedAngularSpectrumAmplitudeStratified} was computed as a Riemann sum with the same $\Delta z$.
Finally, the time step was chosen to satisfy the CFL condition for the frequencies of interest, and ensure stability of the simulations.\cite{treeby_k-wave:_2010}
In all cases, a perfectly matched layer of at least two wavelengths was used at all boundaries (with the number of points chosen to improve the simulation efficiency).
As the ASA is relatively insensitive to noise,\cite{arvanitis_passive_2017} reconstructions did not generally include added noise.
To confirm that this effect persisted with the proposed correction, select reconstructions were performed with added white noise with amplitudes up to 5 times the peak signal level. 

In the biomedical-scale applications, a mean sound speed of \SI{1540}{m/s} was augmented by a \SI{25}{\percent} Gaussian profile with variance of \SI{30}{mm}.
A mean density of \SI{1043}{kg/m\cubed} and attenuation of \SI{0.54}{dB/cm/MHz} were defined for the entire medium\cite{culjat_review_2010} and 99 narrowband sources of \SI{1}{MHz} were distributed uniformly in the medium as in \cref{fig:Biomedical}(a); the simulation grid had resolution $\Delta x = \Delta z = \SI{0.2}{mm}$, and $\Delta t = \SI{50}{ns}$.
For underwater simulations, 70 narrowband sources of  \SI{1}{kHz} were distributed uniformly underwater as shown in \cref{fig:Underwater}(a), and the length of the array located on the water surface was \SI{190}{m}.
The sound speed profile was taken from Ref.~\citenum{munk_ocean_1995} at \ang{0.5}\,N and \ang{100.5}\,W, with attenuation defined from an empirical frequency-dependent model;\cite{jensen_computational_2011}
the simulation grid for the underwater simulations had resolution $\Delta x = \Delta z = \SI{60}{cm}$, and $\Delta t = \SI{5}{\micro\second}$.
Lastly, for the atmospheric simulations, 209 narrowband sources of \SI{1}{Hz} were distributed uniformly above ground as shown in \cref{fig:Atmospheric}(a), and the length of the array located on ground was \SI{6}{km}.
The medium density and sound speeds were defined from the 1972 US Standard Atmosphere,\cite{administration_us_1976} and the altitude- and frequency-dependent attenuation were taken from Ref.~\citenum{sutherland_atmospheric_2004}.
In the atmospheric simulations, the simulation grid had resolution $\Delta x = \Delta z = \SI{10}{m}$, and $\Delta t = \SI{20}{ms}$.


\section{Results}
\label{sec:Results}
\begin{figure}[!ht]
	\centering
    \includegraphics[width=0.8\textwidth]{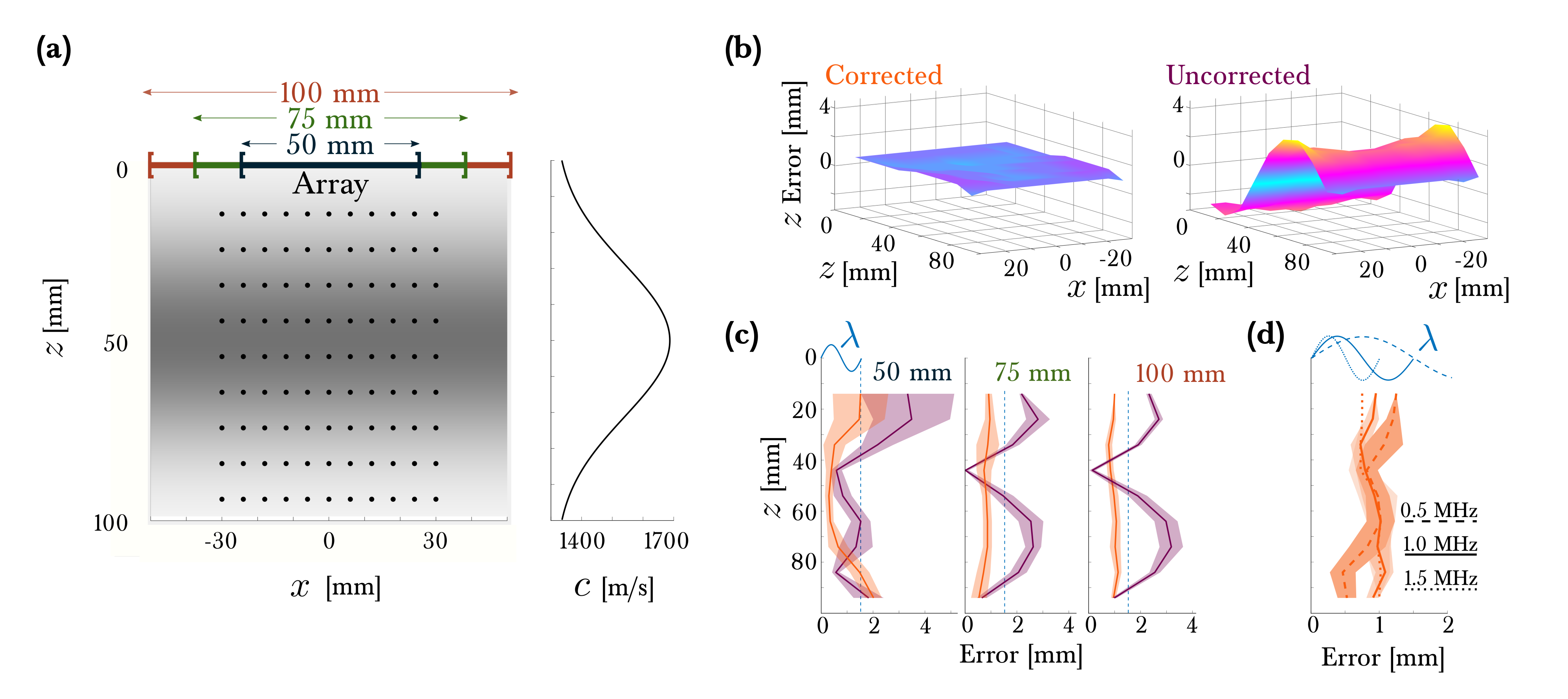}
    \caption{\small Source Localization Improvement with Phase Correction %
    \textbf{(a)}~Arrangement of sources (black circles) relative to the virtual sensor array and stratified sound speed (grayscale and plot at right). %
    Note $z$ direction is downward. %
    \textbf{(b)}~Axial error error with the phase corrections (left) and without (right). %
    \textbf{(c)}~Mean (line) and standard deviation (shaded region) of the errors at each depth $z$ averaged over all transverse positions $x$ for the corrected and uncorrected cases with the indicated aperture. 
    For reference the wavelength of the \SI{1}{MHz} signal is indicated at the dotted line. %
    \textbf{(d)}~Depth dependent errors for a \SI{100}{mm} aperture for the indicated center frequencies, compared with the associated wavelengths.}
    \label{fig:Biomedical}
\end{figure}

\Cref{fig:Biomedical} demonstrates the improvement in source localization due to the phase correction for \SI{1}{MHz} sources.
Without the phase correction, the error was \SI{2.05\pm1.0}{mm}, while with the phase correction it was \SI{0.97\pm0.2}{mm}.
The error in both the corrected and uncorrected case was principally in the vertical (axial) direction ( $|\epsilon_{z,\mathrm{avg}}/\epsilon_{x,\mathrm{avg}}| = 34.6$ and 35.6 for the corrected and uncorrected cases, respectively) and is plotted in \cref{fig:Biomedical}(b).
To understand the effect of the beamforming aperture, we beamformed and localized using three different aperture sizes. 
For the $\SI{50}{mm}$ aperture, some localizations in the corrected case have larger errors, however for larger apertures (\SI{75}{mm} and \SI{100}{mm}) which cover the horizontal extent of the sources, the corrected localization accuracy was within approximately one half wavelength [\cref{fig:Biomedical}(c), orange].
In the uncorrected case, (purple) the error was largest for depths near \SI{25}{mm} and \SI{75}{mm}.
These are the extrema of $\tdi{c}{z}$, and thus where discounting the medium variation is most egregious.

Finally, absolute the localization accuracy did not depend strongly on the wavelength [\cref{fig:Biomedical}(d)].
The absolute localization error was similar for all frequencies (mean \SI{0.91}{mm}, \SI{0.97}{mm}, and \SI{0.88}{mm} for \SI{0.5}{MHz}, \SI{1.0}{MHz}, and \SI{1.5}{MHz}, respectively), as was the improvement relative to the uncorrected case at that frequency (mean improvement \SI{55}{\percent}, \SI{57}{\percent}, and \SI{53}{\percent}).
The short wavelength criterion [\cref{eqn:ShortWavelengthCriterion}] was roughly met in all cases; for the longest wavelength (\SI{500}{kHz}), $|\ttdi{A}{z}| \sim 0.4$.

\begin{figure}[!ht]
    \centering
    \includegraphics[width=0.75\textwidth]{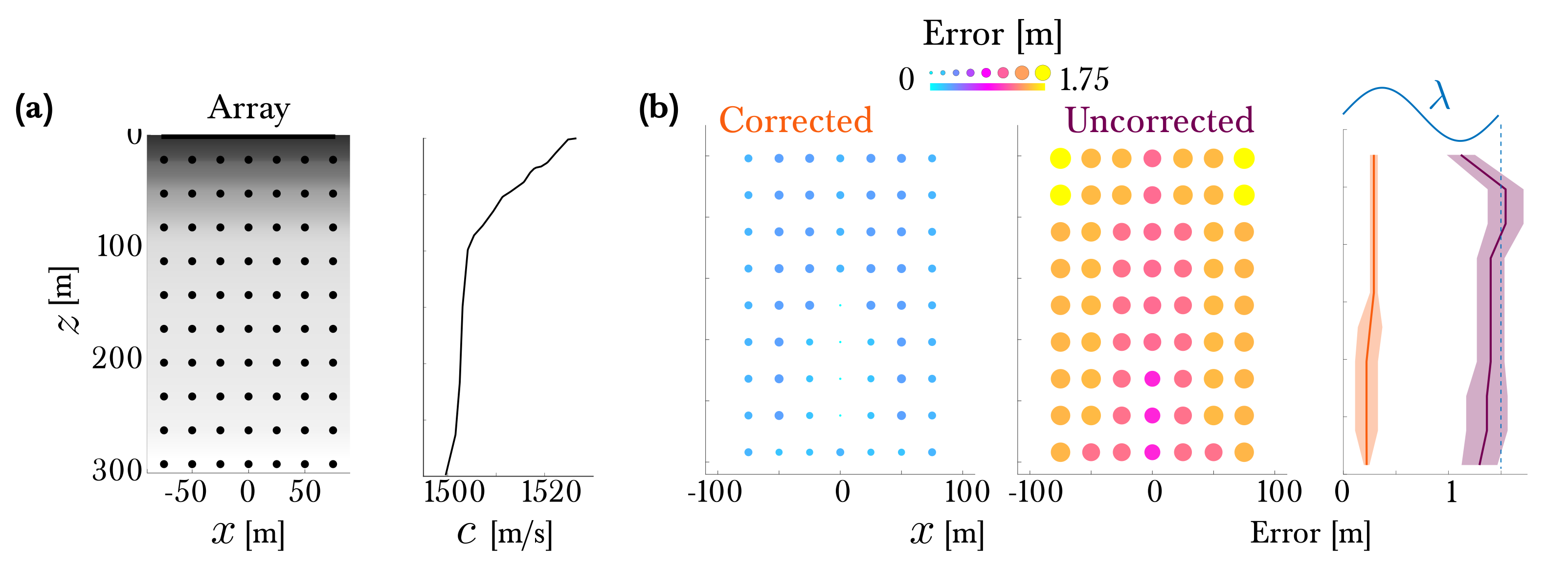}
    \caption{\small Underwater Scaling %
    \textbf{(a)}~Arrangement of sources (black circles) relative to the virtual sensor array and stratified sound speed (grayscale and plot at right). %
    \textbf{(b)}~Total radial error for the corrected (left) and uncorrected (center) cases. 
    At right are shown the mean (line) and standard deviation (shaded region) of the errors at each depth $z$ averaged over all transverse positions $x$ for the corrected and uncorrected cases with the indicated aperture. 
    For reference the wavelength of the \SI{1}{kHz} signal is indicated at the dotted line. }
    \label{fig:Underwater}
\end{figure}

For the simulations with data from measured underwater sound speed profile, the magnitude of the $|c - c_{0}|/c_{0}$ was small---approximately \SI{2}{\percent} [\cref{fig:Underwater}(a)].
Accordingly, the source localization even in the uncorrected case was reasonable, with errors on the order of a wavelength (\SI{1.38\pm0.12}{m} compared with the wavelength of \SI{1.5}{m}) over all depths [\cref{fig:Underwater}(b)].
The correction successfully reduced the error by approximately 5-fold, to \SI{0.26\pm0.08}{m}.
The axial component comprised most of the total error in the uncorrected case ($|\epsilon_{z,\mathrm{avg}}/\epsilon_{x,\mathrm{avg}}| = 9.2$), however in the corrected case, the axial and transverse errors were comparable ($|\epsilon_{z,\mathrm{avg}}/\epsilon_{x,\mathrm{avg}}| = 1.5$)

\begin{figure}[!ht]
    \centering
    \includegraphics[width=0.75\textwidth]{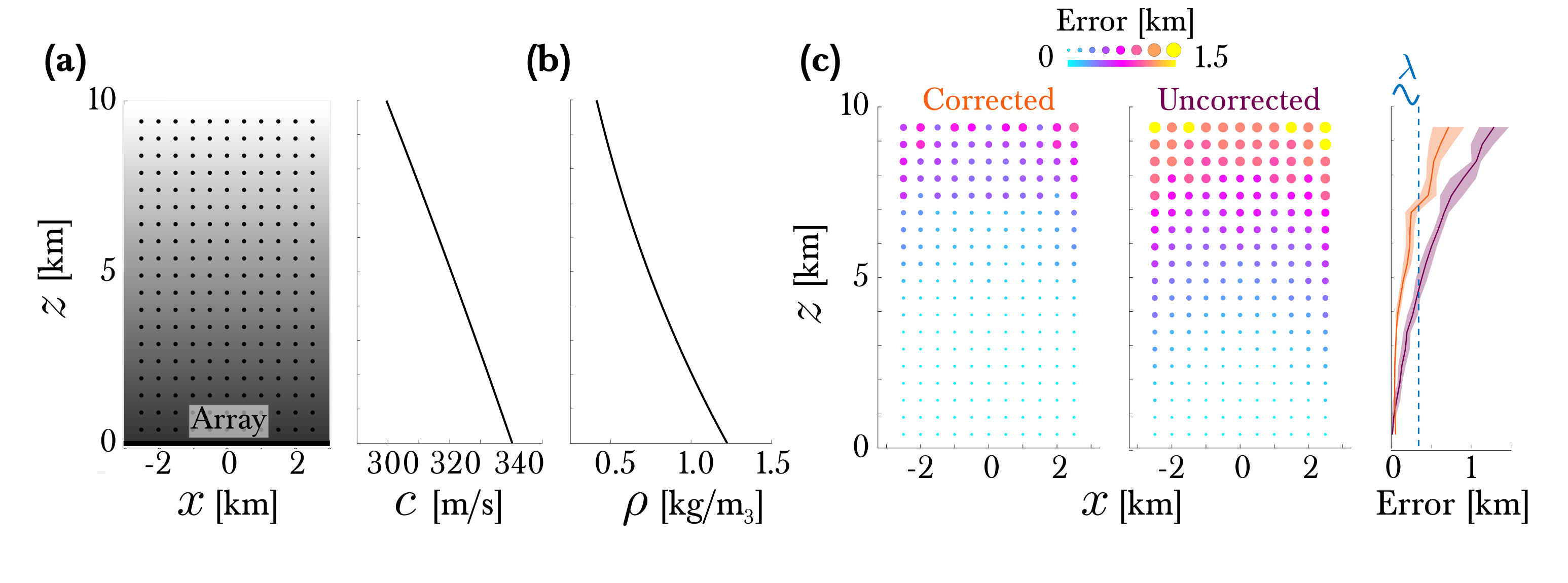}
    \caption{\small Atmospheric Scaling %
    \textbf{(a)}~Arrangement of sources (black circles) relative to the virtual sensor array and stratified sound speed (grayscale and plot at right; note $z$-direction is upward in this figure). %
    \textbf{(b)}~Vertically-dependent medium density used in the simulations.
    \textbf{(c)}~Total radial error for the corrected (left) and uncorrected (center) cases. 
    At right are shown the mean (line) and standard deviation (shaded region) of the errors at each depth $z$ averaged over all transverse positions $x$ for the corrected (orange) and uncorrected (purple) cases with the indicated aperture. 
    For reference the wavelength of the \SI{1}{Hz} signal is indicated at the dotted line. }
    \label{fig:Atmospheric}
\end{figure}

In the atmospheric case, the sound speed decreased approximately linearly over the altitudes considered [up to \SI{9}{km}, \cref{fig:Atmospheric}(a)].
Unlike the other cases considered here, the simulation medium also had significant density variations [$|\rho - \rho_{0}|/\rho_{0}\sim0.5$, \cref{fig:Atmospheric}(b)].
Over the range of positions considered, the mean source localization error was reduced from \SI{474.4\pm395.6}{m} in the uncorrected case to \SI{219.4\pm226.3}{m} with the correction, compared with a wavelength of \SI{343}{m} [\cref{fig:Atmospheric}(c)].

\subsection{Effect of Noise}
\label{sec:Noise}
\begin{figure}
    \centering
    \includegraphics[width=0.7\textwidth]{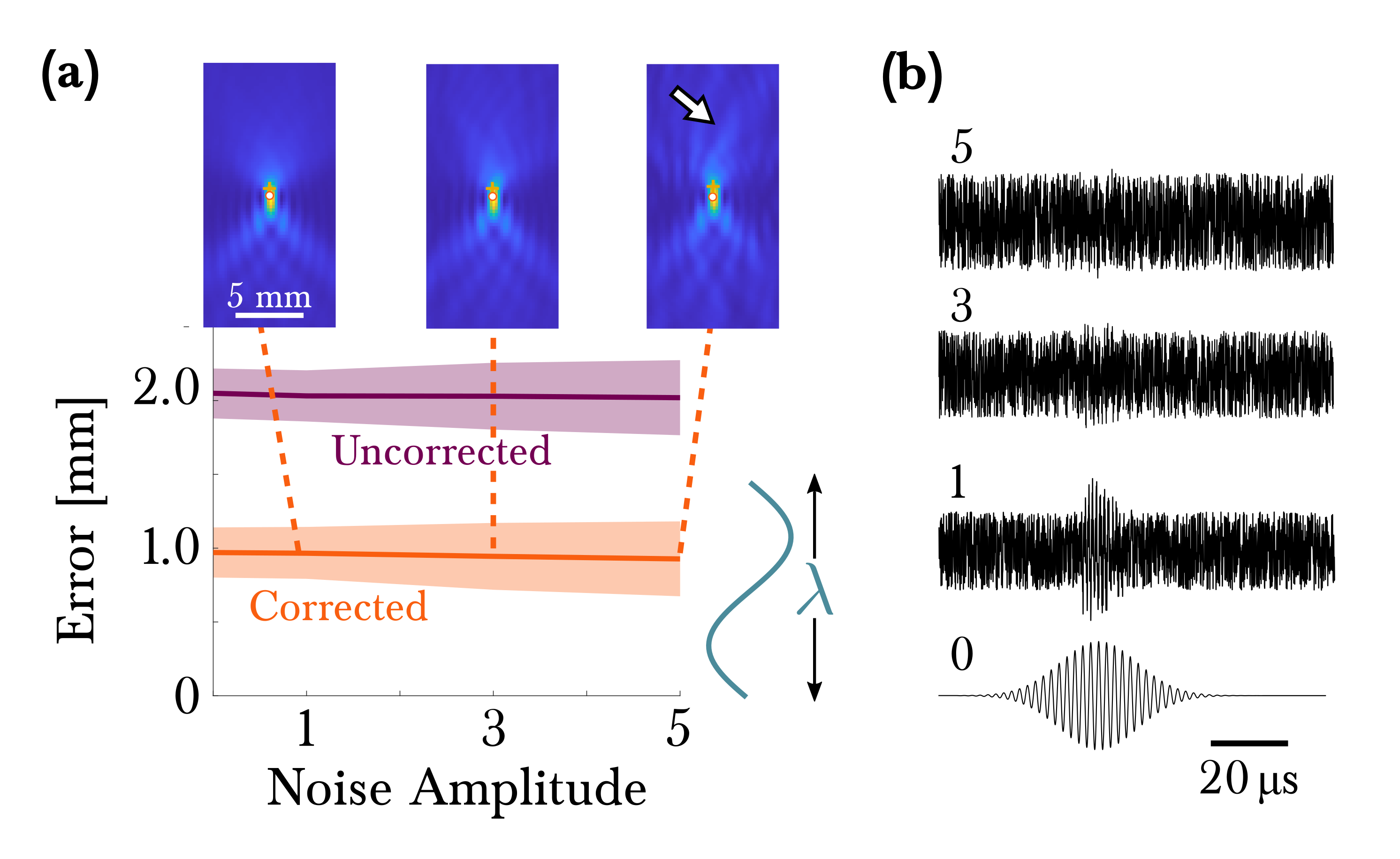}
    \caption{Effect of noise on localization accuracy at the biomedical scale (\cref{fig:Biomedical}, \SI{1}{MHz} sources, \SI{100}{mm} aperture). %
    \textbf{(a)}~Mean localization error as a function of the added noise level for the uncorrected (purple) and corrected (orange) ASA.
    \textbf{(b)}~Center channel waveforms from the simulation after addition of the indicated amount of noise.}
    \label{fig:EffectOfNoise}
\end{figure}

To demonstrate the robustness of both the homogeneous and stratified ASA solutions to noise, the RF data from the simulations whose results are reported in \cref{fig:Biomedical} (\SI{100}{mm} aperture, \SI{1}{MHz} sources) were beamformed after the addition of uniform white noise with amplitude 1 to 5 times that of the maximum value of the recorded RF data.
The resulting PAMs show clear peaks with comparable error to the noiseless case [\cref{fig:EffectOfNoise}(a)] despite apparent total corruption of the time series data [\cref{fig:EffectOfNoise}(b)].
Averaged over the entire grid [\cref{fig:Biomedical}(a)], the error without noise was \SI{0.97 \pm 0.20}{mm} and \SI{2.0 \pm 0.97}{mm} in the corrected and uncorrected cases, respectively. 
At the noise level of 5, the average errors were \SI{0.93 \pm 0.25}{mm} with the correction and \SI{2.0 \pm 1.0}{mm} uncorrected.
The noise manifests in the PAMs as irregular interference patterns [arrow in \cref{fig:EffectOfNoise}(a)], which can cause aberrant high intensities and thus spurious localizations at still larger noise values.
However, \cref{fig:EffectOfNoise} confirms that both the conventional and stratified medium ASA techniques are suitable for noisy conditions.

\subsection{Computational Efficiency}
\label{sec:Efficiency}
The time required for evaluation of \cref{eqn:IntensityMapEquation} depends on the number of frequency bins, signal duration, and the size of the computational grid.
For the cases described here, reconstruction times were on the order of tens of microseconds for the uncorrected case (three frequency bins) and on the order of \SI{100}{ms} when the correction was used.
In all cases, this reduced to tens to a few hundred nanoseconds per pixel in the reconstructed image; see \cref{tab:ComputationTime}
\begin{table}[!htb]
  \renewcommand{\arraystretch}{1.2}
  \caption{\small Mean and standard deviation of the single frequency ASA PAM computation times for uncorrected beamforming [\cref{eqn:AngularSpectrumTransferFunction}], stratified medium corrections [\cref{eqn:LinearizedAngularSpectrumAmplitudeStratified}]
  for each environment.
  The grid size comprises the number of sensors $N_{x}$, number of depth steps $N_{z}$, and number of time steps $N_{t}$.
  }
  \label{tab:ComputationTime}
  \centering
  \begin{tabular}{c c c c c}
    \textbf{Environment} & \textbf{Uncorrected~[ms]} & \textbf{Corrected~[ms]} & \textbf{Grid Size [$N_{x} \times N_{z} \times N_{t}$]} 
    \\ \hline
    Biomedical & $43.4 \pm 43.5$ & $124.7 \pm 125.0$ & $588 \times 600 \times 2000$ 
    \\
    Underwater & $150.5 \pm 2.2$ & $437.5 \pm 12.9$ & $314 \times 320 \times 7000$ 
    \\
    Atmospheric & $40.4 \pm 1.3$ & $112.3 \pm 5.0$ & $588 \times 600 \times 2500$
  \end{tabular}
\end{table}


\section{Discussion}
\label{sec:Discussion}
In this paper, we have presented an analytical phase correction for ASA beamforming in a stratified medium.
Under the assumptions that the sound speed changes slowly compared to a wavelength and that the magnitude of the change is small, an analytical solution the governing equation accurate to first order may be obtained.
Beamforming of passively collected data with the correction mitigates aberration caused by the constant sound speed assumption of the canonical ASA (\cref{fig:PhaseCorrectionConcept}).
Through simulations of point source radiation at biomedical (\cref{fig:Biomedical}), underwater (\cref{fig:Underwater}), and atmospheric (\cref{fig:Atmospheric}) scalings, the error was reduced by \SI{62.6}{\percent}, averaged across all cases. 
Collectively, the results indicate that the method has value for passive source localization at several realistic environments where such information is of considerable interest.\cite{huang_near-field_1991}

While the computational cost was approximately three-fold higher in computational time (mean \SI{147}{ms}), this is still orders of magnitude more efficient than time-domain methods that can handle such corrections natively.
The computation time could be reduced further by reducing the number of receivers (provided the spacing between remains smaller than a half wavelength i.e., the spatial Nyquist frequency),\cite{williams_fourier_1999}
reducing the signal duration or sampling rate $f_{s}$ (provided that the full source signal is captured and $f_{s} \geq 2f_{\mathrm{max}}$, i.e., the Nyquist limit), or by use of parallel computations to reconstruct the maps at each frequency.
Thus the localization method extends the real-time capabilities of the ASA.

The method has a few limitations.
First, the derivations require that the relative changes in sound speed occur over long scales (i.e., $|(\tdi{c}{z})\Delta z/c|$ is very small over a wavelength), and secondarily that these changes relatively small magnitudes (i.e., $|c/c_{0}| \sim 1$).
Thus discrete impedance changes (e.g., layers) violate these assumptions (however such cases may be treated with a layer-by-layer approach\cite{clement_forward_2003}).
Second, only forward propagation is included, such that reflections (e.g., from the ground or water surface) are not accounted for by the method.


\section{Conclusion}
\label{sec:Conclusion}
Here, we derived an analytical, first-order correction to the angular spectrum approach for media with stratified material properties.
This method extends the ability of the in inherently efficient method to localize individual sources at subwavelength accuracy with little additional computational burden.

\section*{Acknowledgements}
\label{sec:Acknowledgements}
Work supported by NIH Grant R00EB016971 (NIBIB) and NSF Grant 1830577 (LEAP HI).

\bibliographystyle{IEEEtran6}
\bibliography{ms}

\end{document}